\documentclass[conference]{IEEEtran}
\IEEEoverridecommandlockouts
\usepackage{amsmath,amssymb,amsfonts}
\usepackage{mathptmx}
\usepackage{algorithmic}
\usepackage{graphicx}
\usepackage[numbers,sort&compress]{natbib}
\usepackage{hyperref}
\makeatletter
\let\old@ps@headings\ps@headings
\let\old@ps@IEEEtitlepagestyle\ps@IEEEtitlepagestyle
\def\confheader#1{%
\def\ps@headings{%
\old@ps@headings%
\def\@oddhead{\strut\hfill#1\hfill\strut}%
\def\@evenhead{\strut\hfill#1\hfill\strut}%
}%
\def\ps@IEEEtitlepagestyle{%
\old@ps@IEEEtitlepagestyle%
\def\@oddhead{\strut\hfill#1\hfill\strut}%
\def\@evenhead{\strut\hfill#1\hfill\strut}%
}%
\ps@headings%
}
\makeatother

\confheader{%
4th International Conference on Electrical Energy Systems (ICEES), Feb. 7-9, 2018, SSNCE, Chennai, TN, INDIA
}

\usepackage[pscoord]{eso-pic}
\newcommand{\placetextbox}[3]{
\setbox0=\hbox{#3}
\AddToShipoutPictureFG{ \put(\LenToUnit{#1\paperwidth},\LenToUnit{#2\paperheight}){\vtop{{\null}\makebox[0pt][c]{#3}}}
}
}
\placetextbox{.2}{0.055}{ 978-1-5386-3695-4/18/\$31.00~\copyright 2018 IEEE}
\begin{document}

\title{Performance analysis of FSO using relays and spatial diversity under log-normal fading channel}

\author{\IEEEauthorblockN{Pranav Kumar Jha$^{*}$, Nitin Kachare, K. Kalyani and {D. Sriram Kumar}}
\IEEEauthorblockA{\textit{Department of Electronics and Communication Engineering},\\ {National Institute of Technology, Tiruchirappalli, Tamil Nadu 620 015, India} \\
$^{*}$jha\_k.pranav@live.com}
}

\maketitle
\begin{abstract}
The performance analysis of free space optical communication (FSO) system using relays and spatial diversity at the source is studied in this paper. The effects of atmospheric turbulence and attenuation, caused by different weather conditions and geometric losses, have also been considered for analysis. The exact closed-form expressions are presented for bit error rate (BER) of M-ary quadrature amplitude modulation (M-QAM) technique for multi-hop multiple-input and single-output (MISO) FSO system under log-normal fading channel. Furthermore, the link performance of multi-hop MISO and multi-hop single-input and single-output (SISO) FSO systems are compared to the different systems using on-off keying (OOK), repetition codes (RCs) and M-ary pulse amplitude modulation (M-PAM) techniques. A significant performance enhancement in terms of BER analysis and SNR gains is shown for multi-hop MISO and multi-hop SISO FSO systems with M-QAM over other existing systems with different modulation schemes. Moreover, Monte-Carlo simulations are used to validate the accuracy and consistency of the derived analytical results. Numerical results show that M-QAM modulated multi-hop MISO and multi-hop SISO FSO system with relays and spatial diversity outperforms other systems while having the same spectral efficiency for each system.
\end{abstract}


\IEEEpeerreviewmaketitle

\section{Introduction}
Free space optical communication (FSO) is a rapidly evolving technology to handle high data rate and has very huge information handling capacity. FSO systems are proven an alternative to the fiber optics technology. In many cases, only the free space transmission is possible to establish a connection between the source and the destination over a line-of-sight (LOS) path.
Optical signal transmission in free space has explored the unexplored areas in wireless communications for decades and appearing as the key of many systems for varieties of applications, such as radio frequency (RF) wireless transmission, satellite communications, long-haul connections and optical fiber back up, etc. FSO systems have to face many challenges, mainly atmospheric turbulence, beam wandering, beam attenuation, weather attenuation, geometric losses and scintillation \cite{navidpour2007ber}. 
\subsection{Related Works}
In literature, different types of channel models have been proposed to exactly model the atmospheric turbulence which matches well with the experimental results. Log-normal, gamma-gamma and negative exponential channel models are used for weak, strong-moderate and strong turbulence conditions, respectively \cite{luong2013effect,rodrigues2013evaluation,peppas2012simple}. Fading and attenuation has affected the performance of FSO systems in large extent. Turbulence and weather conditions, such as snow, fog and haze also affecting the FSO transmission. Further, geometric loss has also lessened the performance of FSO systems. In previous works, varieties of spatial diversity techniques have been proposed to mitigate the effect of atmospheric turbulence in FSO communications \cite{abaza2014diversity,tsiftsis2009optical} where three primary spatial diversity schemes are more popular among the available diversity schemes, such as orthogonal space time block codes (OSTBCs), transmit laser selection (TLS) and repetition codes (RCs). Here, TLS is considered as the best spatial diversity scheme with channel state information (CSI) at the source end, but at the cost of increased system complexity \cite{garcia2009selection,safari2008relay}. With moderate system complexity in different channel conditions, RCs outperforms its counterpart OSTBCs. Apart from spatial diversity schemes, some other techniques  have also been proposed to mitigate turbulence effects such as relay assisted techniques \cite{safari2008relay}, error-correcting codes \cite{abaza2011ber,zhu2002free} and maximum likelihood estimation \cite{zhu2000maximum}, out of which relay assisted technique appears to be prominent as several short links are used instead of a long communication link for efficient data transmission over FSO channels. 
\subsection{Motivation and Contribution}
In previous works, most of the systems introduced are using pulse amplitude modulation (PAM) as a high level modulation technique, but very few have considered to use quadrature amplitude modulation (QAM) scheme, which is the motivation behind this work.
In this paper, with the help of decode and forward (DF) relay for intensity modulation and direct detection (IM/DD), M-ary QAM (M-QAM) modulation is used against a log-normal fading channel to study the link performance of different architectures of FSO systems for short distance communication links under different weather conditions. Here, bit error rate (BER) is considered as a performance metric which is used to measure the performance of the FSO system. For analysis purposes, the effect of turbulence, path loss, scattering and scintillation have also been taken into account. Assuming the correlation effects among the transmitters, the performance of multi-hop multi-input and single-output (MISO) and multi-hop single-input and single-output (SISO) system with M-QAM is compared with multi-hop MISO and SISO system using M-PAM, MISO with RCs and OOK and SISO with on-off keying (OOK) \cite{abaza2015performance}. In all cases, it is assumed that the spectral efficiency is same. At the receiver, maximum likelihood (ML) decoding technique is employed for receiving the M-QAM signals.
\subsection{Organization}
The rest of the paper is organized as follows: in Section \ref{sec2}, the DF relay FSO system is described for multi-hop MISO configurations under various modulation techniques and the mathematical expressions of BER have been presented for M-QAM and $\textrm{M}^2$-QAM modulation schemes. Section \ref{sec3} presents the numerical results where the parameters, considered for simulation purposes, are tabulated in Table \ref{tab:1}. Section \ref{sec4} concludes the work.
\section{System Description} 
\label{sec2}
\subsection{Multi-hop DF relay System}
\subsubsection{M-QAM Transmission}
\begin{figure}
\includegraphics[width=\linewidth]{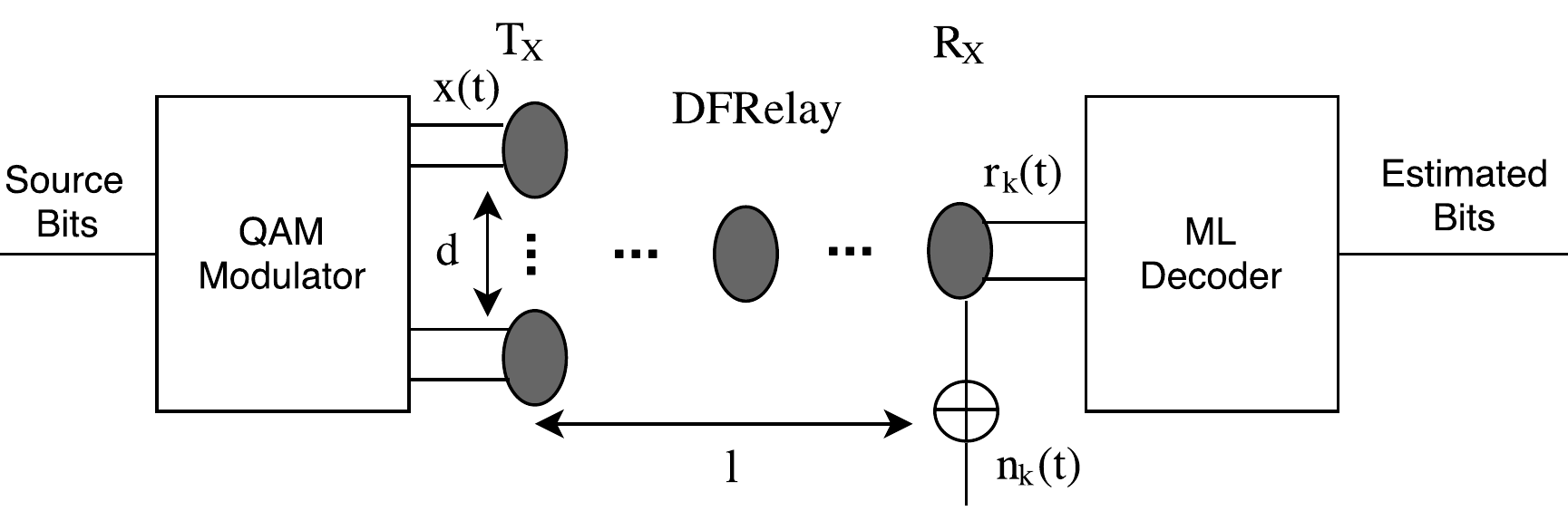}
\caption{A MISO multi-hop DF relay FSO System, which is not benefited by the use of transmit diversity.}
\label{fig:3}
\end{figure}

For transmission of signals, a MISO system is considered with M transmitters in the form of lasers and a photo detector as receiver, which is shown in Fig. \ref{fig:3}. At the source, transmission of signals is in the form of binary data, modulated by a $M_I \times M_Q$ electrical-QAM modulator. For the transmission through M paths, mapping of data symbols into QAM symbols is done with the help of Gray coding. In addition, to mitigate the spatial correlation effect, sufficient channel distance is considered.

The probability of error of QAM signals can be easily calculated from the probability of error of PAM signals as the phase-quadrature signal components can be perfectly extracted at the QAM demodulator. For the M-ary QAM signals, the probability of a correct decision is written as \cite{proakis2001digital}
\begin{equation}\label{eq22}
P_C= (1-P_{\sqrt M} )^2,
\end{equation}
where $P_{\sqrt M}$ is the probability of error for $\sqrt {M}$-ary PAM with half the average power in each quadrature component of the signal of the corresponding QAM system. The probability of error of M-ary PAM can be written as
\begin{equation}\label{eq23}
P_{\sqrt M}=2\bigg(1-\frac{1}{\sqrt M} \bigg)Q\bigg(\frac{3log_2 M\gamma}{M-1}\bigg),
\end{equation}
where $\gamma$ represents the average SNR per symbol. Consequently, the probability of a symbol error for M-ary QAM signals can be written as
\begin{equation}\label{eq24}
P_M=1-(1-P_{\sqrt M})^2,
\end{equation}
which is valid for $M=2^k$ and even $k$. By using the optimum detector, the tightly upper bounded symbol error probability can be written as
\begin{equation}\label{eq25}
\begin{split}
P_M\leq 1-\bigg[1-2Q\bigg(\sqrt{\frac{3 \gamma}{M-1}}\bigg)\bigg]\\
\leq 4Q\bigg(\sqrt{\frac{3 log_2{M}\gamma}{M-1}}\bigg).
\end{split}
\end{equation}
Here, introduction of relays creates some effect in FSO system for SISO and MISO configurations with the help of DF relays, which helps to achieve significant improvement in the link performance. For the system under consideration, $(K-1)$ DF relays are set up in between the source and the destination for $K$ hops. The received signal at the $k$th hop can be written as 
\begin{equation}\label{eq26}
r_k(t) = x(t)\eta I_k + n_k(t), \ k=1,2,\ldots K.
\end{equation}
In this scheme, $K$ time slots are required for transmission of signals from source to destination, unlike MISO and SISO systems. For the benefit of achieving similar spectral efficiency, $2^K$-ary M-QAM modulation technique is employed. DF relays used in between the transceiver pairs create shorter communication links which help to reduce the effects of turbulence and path losses,  significantly. The upper bound BER for multi-hop DF relay system can be written as \cite{abaza2015performance}
\begin{equation}\label{eq27}
\textrm{BER} \leq 1-\prod_{k=1}^{K}(1-\textrm{BER}_k).
\end{equation}
As the identical statistical properties are considered for all hops, the above expression (\ref{eq27}) can be approximated by  
\begin{equation}\label{eq28}
\textrm{BER} \approx \frac{1}{2}[1-(1-2\textrm{BER}_k )^K].
\end{equation}
The conditional BEP of M-QAM is given as \cite{singh4performance}
\begin{equation}\label{eq29}
P(e|\gamma)\approx \frac{2\Big(1-\frac{1}{\sqrt M}\Big)}{\textrm{log}_2 M}\Bigg[Q\Bigg(\sqrt{\frac{3\textrm{log}_2M \gamma}{2(M-1)}} \Bigg)\Bigg],
\end{equation}
and the instantaneous SNR $\gamma$ is given by 
\begin{equation}\label{eq30}
\gamma = \frac{2P^2}{\sigma_n^2 R},
\end{equation}
where $P$ implies the signal power and $R$ signifies the bit rate. 
Now, by applying the approximate $Q$ function \citep{abaza2015performance} on (\ref{eq29}), the conditional BEP of M-QAM can be written as
\begin{equation}\label{eq31}
\begin{split} 
P(e|\gamma) & \approx \frac{2\Big(1-\frac{1}{\sqrt M}\Big)}{\textrm{log}_2 M}\\
& \times\bigg[\frac{1}{12}\textrm{exp}\bigg(-\frac{3 \textrm{log}_2 M \gamma}{4(M-1)}\bigg) + \frac{1}{4}\textrm{exp}\bigg(-\frac{ \textrm{log}_2 M \gamma}{(M-1)}\bigg)\bigg],
\end{split}
\end{equation}
which is the general expression of the conditional BEP for FSO system using M-QAM. Now, the BER expression of the $k$th hop can be derived using Hermite polynomial  and $Q$-function approximation as in \cite{abaza2015performance}. The closed-form expression of BER for M-QAM modulation can be written as \cite{abaza2015performance}
\begin{equation}\label{eq32}
\begin{split}
\textrm{BER}_k & \approx \\ & \frac{G}{12}\sum_{i=1}^{N}w_i \textrm{exp}\bigg(-\frac{3\textrm{log}_2M \beta_{kn}^2 \overline\gamma e^{-4\sigma_k^2+x_i\sqrt{32\sigma_k^2}}}{4(M-1)}\bigg)
\\
& +\frac{G}{4}\sum_{i=1}^{N}w_i \textrm{exp}\bigg(-\frac{\textrm{log}_2M \beta_{kn}^2 \overline\gamma e^{-4\sigma_k^2+x_i\sqrt{32\sigma_k^2}}}{(M-1)}\bigg),
\end{split}
\end{equation}
where $G =2\big(1-\frac{1}{\sqrt M}\big)/( \textrm{log}_2 (M) \sqrt{\pi})$, $\beta_{kn}$ represents the normalized path loss coefficient, $x_i$ and $w_i$ are the zeros and the weights of the Hermite polynomial, $\sigma_k^2$ is the channel variance and $\overline\gamma$ is the average SNR. Substituting (\ref{eq32}) in (\ref{eq27}) and (\ref{eq28}), the upper bound and the average BER  expressions of multi-hop system can be obtained, respectively. The BER expressions of M-PAM, SISO with OOK, and MISO with RCs and OOK are taken from \citep{abaza2015performance} for analysis and comparison purpose.
\subsubsection{$M^2$-QAM Transmission}
The  $M^2$ symbols of $M^2$-QAM  made up of an in-phase and quadrature phase component basis function, orthogonal to each other. In each symbol duration, the two basis functions are modulated with the independent data resulting a multiplication by a series of M amplitude values to each basis function to comprise the $M^2$ symbols \cite{hranilovic2006wireless}. 
The constellation of QAM shows a two dimensional regular array of points and the minimum spacing between the points is prescribed by the amount of DC bias added by virtue of the non-negativity constraint, which is \cite{hranilovic2006wireless}
\begin{equation*}
d_{min}=\frac{P}{M-1}\sqrt{\frac{2\textrm{log}_2 M}{R}},
\end{equation*}
where R signifies the bit rate.
The probability of symbol error is estimated by using the union bound approximation \citep{abaza2015performance}. By evaluating the average number of neighbors $d_{min}$ away from every constellation point, $P_{esym}$ is approximated as
\begin{equation*}
P_{esym}=\frac{4M-1}{m}\cdot Q\Bigg(\frac{P}{M-1}\sqrt{\frac{1}{4 R_s \sigma^2}}\Bigg),
\end{equation*}
where $R_s=R/\textrm{log}_2 M^2$.
By applying the Gray coding approximation, the conditional BEP of $M^2$-QAM is given by \cite{hranilovic2006wireless}
\begin{equation}\label{eq33}
P(e|\gamma)\approx \frac{2\Big(M-1\Big)}{M\textrm{log}_2M}\Bigg[Q\Bigg(\sqrt{\frac{\textrm{log}_2M \gamma}{4{(M-1)}^2}}\Bigg)\Bigg],
\end{equation}
where instantaneous SNR 
$
\gamma = \frac{2P^2}{\sigma_n^2 R}.
$
Now, applying the approximate $Q$ function on (\ref{eq33}), the conditional BEP of M-QAM is given as
\begin{equation}\label{eq34}
\begin{split}
P(e|\gamma) & \approx 
\frac{2\Big(M-1\Big)}{M\textrm{log}_2M}\\
& \times \bigg[\frac{1}{12}\textrm{exp}\bigg(-\frac{ \textrm{log}_2 M \gamma}{8(M-1)^2}\bigg) + \frac{1}{4}\textrm{exp}\bigg(-\frac{\textrm{log}_2 M \gamma}{6(M-1)^2}\bigg)\bigg],
\end{split}
\end{equation}
which represents the typical expression of the conditional BEP for the FSO system with $M^2$-QAM. The closed form expression of BER of $M^2$-QAM modulation derived as discussed earlier, can be written as
\begin{equation}\label{eq35}
\begin{split}
\textrm{BER}_k & \approx \frac{G}{12}\sum_{i=1}^{N}w_i \textrm{exp}\bigg(-\frac{\textrm{log}_2M \beta_{kn}^2 \overline\gamma e^{-4\sigma_k^2+x_i\sqrt{32\sigma_k^2}}}{8(M-1)^2}\bigg)
\\
& + \frac{G}{4}\sum_{i=1}^{N}w_i \textrm{exp}\bigg(-\frac{\textrm{log}_2M \beta_{kn}^2 \overline\gamma e^{-4\sigma_k^2+x_i\sqrt{32\sigma_k^2}}}{6(M-1)^2}\bigg),
\end{split}
\end{equation}
where $G = 2\big(M-1\big)/M\textrm{log}_2M \sqrt{\pi}$ and $\beta_{kn}$ is the normalized path loss coefficient with reference to the direct link of the multi-hop system. Substituting (\ref{eq35}) in (\ref{eq27}) and (\ref{eq28}), the upper bound BER expression and the average BER  expression for the multi-hop system can be retrieved, respectively.
\begin{figure}
\includegraphics[width=\linewidth]{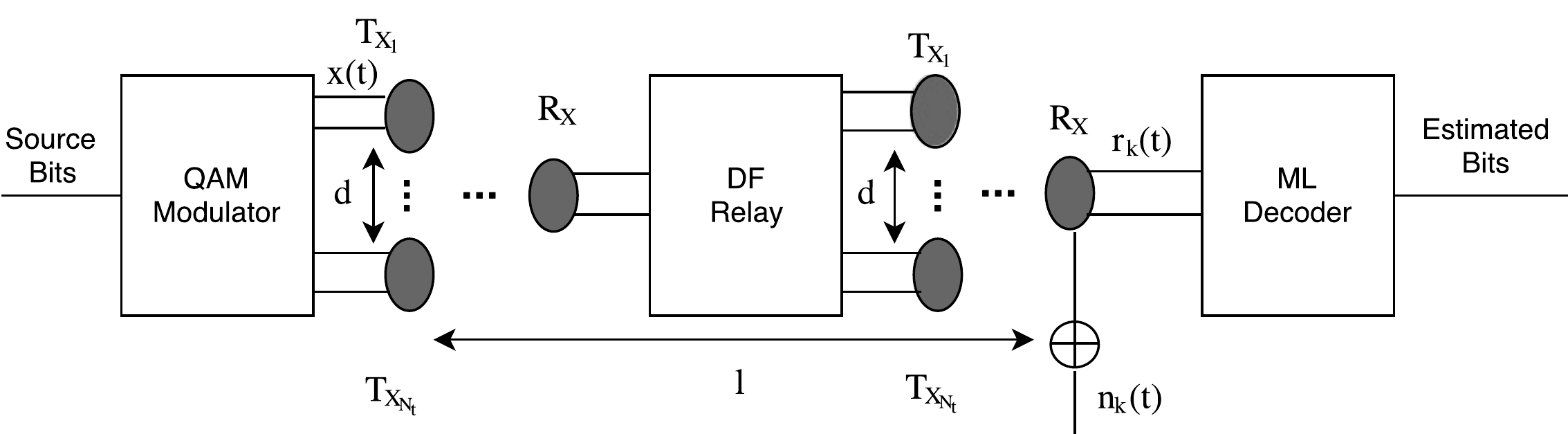}
\caption{A hybrid MISO multi-hop DF relay FSO System which is benefited by relays and transmit diversity, both.}
\label{fig:4}
\end{figure}
\subsection{MISO multi-hop DF relay system}
This is a hybrid technique which uses the advantages of FSO systems with relays and transmit diversity to deteriorate the channel attenuation and turbulence effects. The block diagram of this system model is shown in Fig. \ref{fig:4}.
Here, $K-1$ relays are employed to receive the signals from $N_t$ transmitters which, simultaneously, re-transmit those signals with the help of RCs. For this system, the number of hops are considered identical with the number of transmitters per relay. The signal received at each hop can be given as
\begin{equation}\label{eq36}
r_k (t) = x(t)\eta\sum_{i=1}^{N_t}I_{ki} +n_k (t),  \ k=1,2 \ldots K.
\end{equation}
Similarly, using the conditional BEP expression for M-QAM, the BER expression for $k$th hop can be written as \cite{abaza2015performance}
\begin{equation}\label{eq37}
\begin{split}
\textrm{BER}_k
& \approx \sum_{n_1=1}^{N}\ldots \sum_{n_{N_t}}^{N}\Bigg[\prod_{i=1}^{N_t}\frac{w_{n_i}}{\sqrt{\pi}}\Bigg]\frac{F}{12}\textrm{exp}\\
& \times \Bigg(-\frac{3\textrm{log}_2(M) \beta_{kn}^2\overline\gamma}{4(M-1)N_t}
\sum_{i=1}^{N_t}\Bigg[\textrm{exp}\bigg(\sqrt{32}\sum_{j=1}^{N_t}{c_{ij}^{'}x_{nj}-4\sigma_k^2}\bigg)\Bigg]\Bigg)\\
& + \sum_{n_1=1}^{N}\ldots \sum_{n_{N_t}}^{N}\Bigg[\prod_{i=1}^{N_t}\frac{w_{n_i}}{\sqrt{\pi}}\Bigg]\frac{F}{4}\textrm{exp}\\
& \times \Bigg(-\frac{\textrm{log}_2(M) \beta_{kn}^2\overline\gamma}{(M-1)N_t}
\sum_{i=1}^{N_t}\Bigg[\textrm{exp}\bigg(\sqrt{32}\sum_{j=1}^{N_t}{c_{ij}^{'}x_{nj}-4\sigma_k^2}\bigg)\Bigg]\Bigg),
\end{split}
\end{equation}
and using the conditional BEP equation of $M^2$-QAM, the BER expression for the $k$th hop has been given as
\begin{equation}\label{eq38}
\begin{split}
\textrm{BER}_k & \approx
\sum_{n_1=1}^{N}\ldots \sum_{n_{N_t}}^{N}\Bigg[\prod_{i=1}^{N_t}\frac{w_{n_i}}{\sqrt{\pi}}\Bigg]
 \frac{F}{12}\textrm{exp}\\
& \times\Bigg(-\frac{\textrm{log}_2(M) \beta_{kn}^2\overline\gamma}{8(M-1)N_t}
\sum_{i=1}^{N_t}\Bigg[\textrm{exp}\bigg(\sqrt{32}\sum_{j=1}^{N_t}{c_{ij}^{'}x_{nj}-4\sigma_k^2}\bigg)\Bigg]\Bigg)\\
& + \sum_{n_1=1}^{N}\ldots \sum_{n_{N_t}}^{N}\Bigg[\prod_{i=1}^{N_t}\frac{w_{n_i}}{\sqrt{\pi}}\Bigg]
\frac{F}{4}\textrm{exp}\\
& \times \Bigg(-\frac{\textrm{log}_2(M) \beta_{kn}^2\overline\gamma}{6(M-1)N_t}
\sum_{i=1}^{N_t}\Bigg[\textrm{exp}\bigg(\sqrt{32}\sum_{j=1}^{N_t}{c_{ij}^{'}x_{nj}-4\sigma_k^2}\bigg)\Bigg]\Bigg),
\end{split}
\end{equation}
where $F = 2\big(1-\frac{1}{\sqrt M}\big)/\textrm{log}_2 (M)$ for (\ref{eq37}) and $F=2\big(M-1\big)/M\textrm{log}_2M$ for (\ref{eq38}) and $c_{ij}^{'}$ represents the $(i,j)$th coefficients of the  spatial covariance matrix Γ$\Gamma_{sq}^{'}=\Gamma^{'1/2}$. By substituting these results in (\ref{eq27}) and (\ref{eq28}), expressions of the upper bound BER and the closed-form average BER of the multi-hop MISO system can be obtained, respectively. The BER expressions for other configurations of FSO system are shown in \cite{abaza2015performance}.
\section{Numerical Analysis and Discussions}
\label{sec3}
Numerical results are presented in this section which verify the derived analytical results using Monte-Carlo simulations. For
simulation purposes, a minimum of $10^6$ bits are relayed for respective SNR
values and increases with increasing SNR. The parameters needed to achieve a target BER of $10^{-9}$ are defined and tabulated in Table \ref{tab:1}. Further, all hops are considered equidistant with each other. The scintillation index (SI) is $\leq$ 0.75 for log-normal channel and 
$\sigma_x=\sqrt{\textrm{ln}(\textrm{SI})+1/2}$ \cite{moradi2011switched}. Hence, for SI = 0.75, $\sigma_x\leq0.374$ is required, which represents the maximal value assumed for the analysis purpose. 
\begin{figure}
\includegraphics[width=\linewidth]{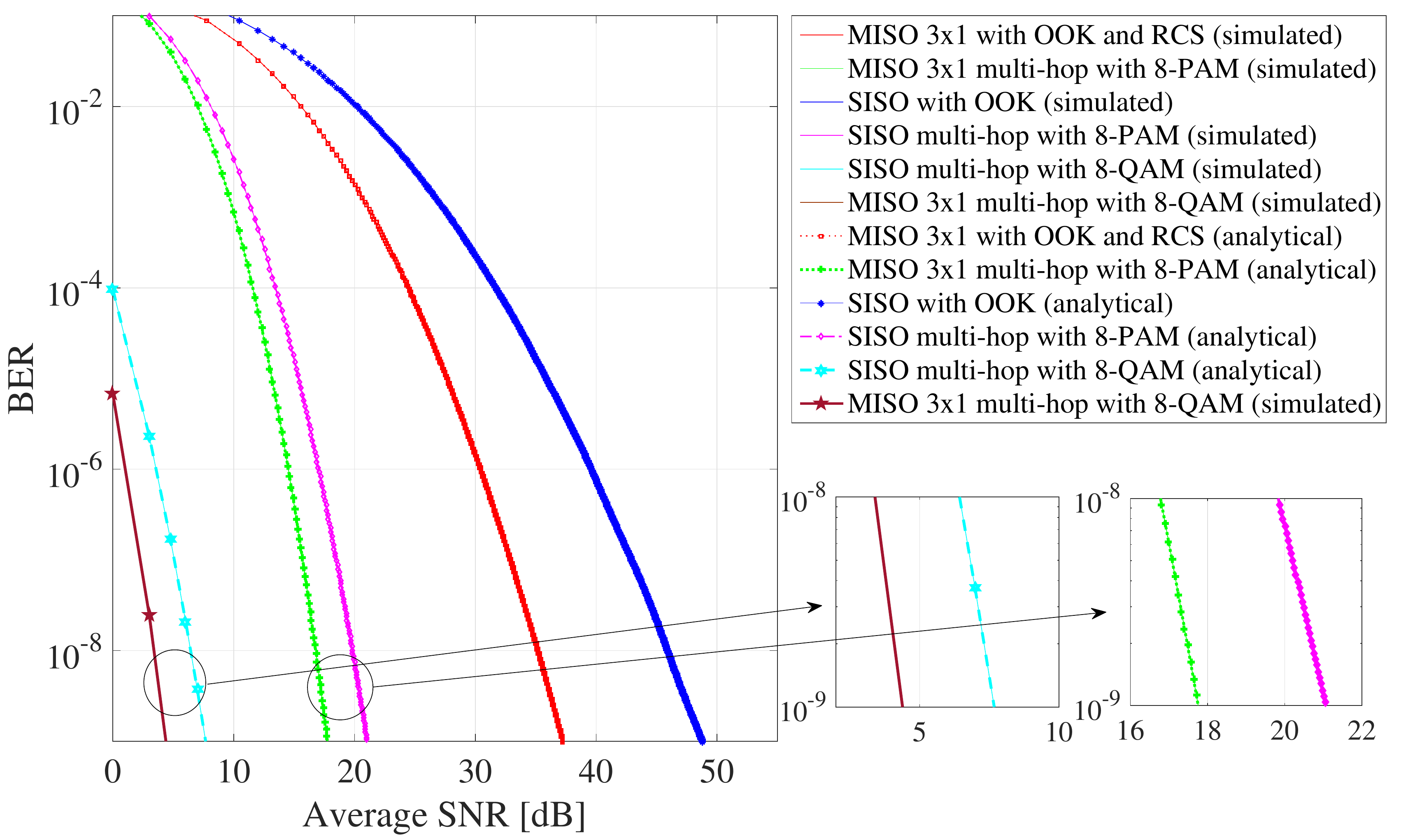}
\caption{BER of multi-hop FSO system with 8-QAM, 8-PAM, SISO-OOK and MISO-RCs and OOK in clear weather condition.}
\label{fig:7}
\end{figure}
\begin{table}[htbp]
\caption{Simulation Parameters}
\begin{center}
\begin{tabular}{|l|l|l|}
\hline
\textbf{FSO Parameters} & \textbf{Numerical Values}\\
\hline
Relay Spacing & 400 m \\
\hline
Wavelength ($\lambda$) & 1550 nm \\
\hline
Link Distance $(l)$ & 1200 m\\
\hline
Beam Divergence Angle $(\theta_T)$ &	2 mrad\\
\hline
Correlation Coefficient $(\rho)$ & 0.3\\
\hline
Tx and Rx Aperture Diameter $(D_R\textrm{ and }D_T)$	&
20 cm  \\
\hline
Attenuation Constant $(\alpha)$ & 0.43 dB/km (Clear Weather)\\&
20 dB/km (Light Fog)\\
\hline
Refractive Index Constant $(C_n^2)$ &
$5\times10^{-14} \textrm{m}^{-(2/3)}$ (Clear Weather)\\&
$1.7\times 10^{-14} \textrm{m}^{-2/3}$
(Light Fog)\\
\hline
\end{tabular}
\end{center}
\label{tab:1}
\end{table}

Fig. \ref{fig:7} represents the BER of multi-hop FSO system, where the number of relays is two and the number of transmit lasers is one for SISO scheme or three for MISO scheme, wherever required with 8-QAM, 8-PAM, MISO with RCs and OOK and SISO-OOK modulation techniques. It demonstrates the effect on the performance gain of MISO and SISO multi-hop FSO systems for different modulation techniques under clear weather conditions. Further, at the target BER, SNR gains of 19.52 dB and 27.65 dB is achieved for multi-hop MISO and multi-hop SISO systems as compared to MISO with OOK and RCs and SISO with OOK, respectively, for 8-PAM modulation scheme, whereas with 8-QAM modulation, SNR gains of 13.21 dB and 13.44 dB is achieved for multi-hop MISO and multi-hop SISO systems over PAM modulated multi-hop MISO and multi-hop SISO systems, respectively, maintaining the same number of transmitters. Furthermore, results show that, multi-hop MISO system outperforms multi-hop SISO system by 3.1 dB and 3.33 dB, respectively, for 8-QAM and 8-PAM. 
\begin{figure}
\includegraphics[width=\linewidth]{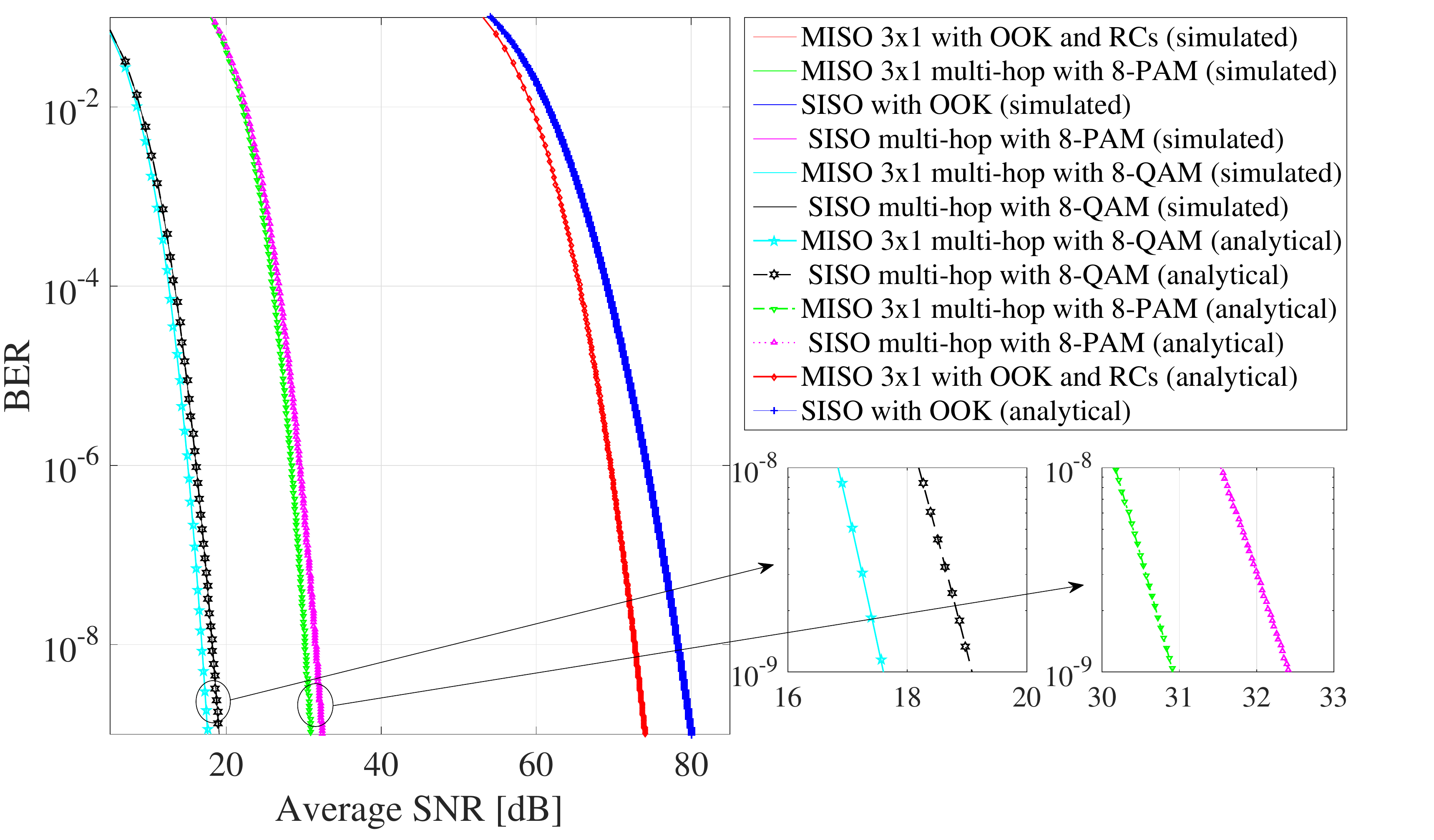}
\caption{BER of multi-hop FSO system with 8-QAM, 8-PAM, SISO-OOK and MISO-RCs and OOK in  light fog condition.}
\label{fig:8}
\end{figure}

In Fig. \ref{fig:8}, BER for multi-hop FSO system with two relays and one or three transmit lasers has been shown under light fog conditions for 8-QAM, 8-PAM, MISO with RCs and OOK and SISO with OOK modulation schemes. For two relay multi-hop system, 8-QAM modulation scheme provides an equal spectral efficiency for SISO and MISO systems. Derived analytical results show that multi-hop system still outperforms other systems and the increment in the performance of the system is significant. For 8-PAM modulation scheme, SNR gains of 43.13 dB and 47.64 dB are achieved at the target BER of multi-hop MISO and multi-hop SISO systems in comparison with MISO with RCs and OOK and SISO-OOK modulated systems, whereas using 8-QAM technique, system outperforms the 8-PAM system with 13.34 dB and 13.66 dB of SNR gains for multi-hop MISO and multi-hop SISO systems, respectively. In addition, 8-QAM and 8-PAM multi-hop MISO systems provide the performance gain of 1.16 dB and 1.48 dB over the 8-QAM and 8-PAM multi-hop SISO system, respectively.
\section{Conclusions}
\label{sec4}
The effects of moderate turbulence under clear weather condition and weak turbulence under light fog conditions on the performance of different types of FSO systems are studied in this paper. It is evident from the derived analytical results that diverse conditions in the atmosphere have a huge effect on the performance of MISO and SISO multi-hop FSO systems. Furthermore, multi-hop MISO system with M-QAM modulation outperforms other systems, such as SISO with OOK, MISO with RCs and OOK and M-PAM multi-hop SISO and MISO systems in terms of SNR gains and BER performance considering the same spectral efficiency for all the systems. Moreover, the results also demonstrate that multi-hop systems are capable of mitigating the effects of turbulence and path losses caused by geometric losses and attenuations, whereas MISO systems can counteract turbulence effects only. Consequently, it can also be stated that the overall performance of the FSO system can be improved significantly by increasing the number of relays where the introduction of spatial diversity could also be of great importance as well.

\bibliographystyle{IEEEtran}
\bibliography{references}

\end{document}